\documentclass[a4paper]{article}

\usepackage[colorlinks=true,linkcolor=blue,citecolor=blue,filecolor=blue,urlcolor=blue]{hyperref}
\usepackage{amsmath}

\usepackage{natbib,doi}
\usepackage{graphicx}
\usepackage{placeins}
\usepackage{sidecap}
\usepackage{xspace}

\newcommand{\Dt}{\Delta t}
\newcommand{\ie}{{\it i.e.}\xspace}
\newcommand{\eg}{{\it e.g.}\xspace}

\newcommand{\pa}{\textit{p.a.}\xspace}
\newcommand{\vs}{\textit{vs.}\xspace}
\newcommand{\psa}{$\sqrt{\textit{p.a.}}$}

\newcommand{\be}{\begin{equation}}
\newcommand{\ee}{\end{equation}}
\newcommand{\bea}{\begin{eqnarray}}
\newcommand{\eea}{\end{eqnarray}}

\newcommand{\supm}{^{\text{m}}}
\newcommand{\mum}{\mu^{\text{m}}}
\newcommand{\murm}{\mu_{\text{riskless}}^{\text{m}}}
\newcommand{\muem}{\mu_{\text{excess}}^{\text{m}}}
\newcommand{\sigmam}{\sigma^{\text{m}}}
\newcommand{\lm}{l^{\text{m}}}
\newcommand{\loptm}{l_{\text{opt}}^{\text{m}}}
\newcommand{\loptestm}{\widehat{ l_{\text{opt}}^{\text{m}}}}

\newcommand{\glm}{g_l^{\text{m}}}
\newcommand{\gestm}{\widehat{g^{\text{m}}}}
\newcommand{\gens}{{g}^{\text{m}}_{\ave{}}}
\newcommand{\gtm}{g^{\text{m}}_t}
\newcommand{\pim}{\pi^{\text{m}}} 
\newcommand{\xl}{x_{l}}
\newcommand{\xlc}{x_{l}^{d}}

\newcommand{\supc}{^{\text{d}}}
\newcommand{\muc}{\mu^{\text{d}}}
\newcommand{\murc}{\mu_{\text{riskless}}^{\text{d}}}
\newcommand{\muec}{\mu_{\text{excess}}^{\text{d}}}
\newcommand{\sigmac}{\sigma^{\text{d}}}
\newcommand{\lc}{l^{\text{d}}}
\newcommand{\loptc}{l_{\text{opt}}^{\text{d}}}

\newcommand{\gtc}{g^{\text{d}}_t}
\newcommand{\pic}{\pi^{\text{d}}} 

\newcommand{\supr}{^{\text{r}}}
\newcommand{\mur}{\mu^{\text{r}}}
\newcommand{\murr}{\mu_{\text{riskless}}^{\text{r}}}
\newcommand{\muer}{\mu_{\text{excess}}^{\text{r}}}
\newcommand{\sigmar}{\sigma^{\text{r}}}
\newcommand{\lr}{l^{\text{r}}}
\newcommand{\loptr}{l_{\text{opt}}^{\text{r}}}

\newcommand{\pir}{\pi^{\text{r}}} 

\newcommand{\Ito}{It\^{o}}

\newcommand{\SP}{S\&P500\xspace} 
\newcommand{\SPtick}{SPX\xspace} 

\newcommand{\ave}[1]{\left\langle#1 \right\rangle}

\newcommand{\elabel}[1]{\label{eq:#1}}
\newcommand{\eref}[1]{equation~(\ref{eq:#1})}

\newcommand{\Eref}[1]{Equation~(\ref{eq:#1})}

\newcommand{\flabel}[1]{\label{fig:#1}}
\newcommand{\fref}[1]{figure~\ref{fig:#1}}
\newcommand{\Fref}[1]{Figure~\ref{fig:#1}}

\newcommand{\tlabel}[1]{\label{tab:#1}}
\newcommand{\tref}[1]{table~\ref{tab:#1}}

\begin{document}

\title{Leverage efficiency}
\author{Ole Peters$^{1,2}$\thanks{Email: \texttt{o.peters@lml.org.uk}}~~and~Alexander Adamou$^1$\thanks{Email: \texttt{a.adamou@lml.org.uk}}\\
$^1$London Mathematical Laboratory, 8 Margravine Gardens, London W6 8RH, UK\\
$^2$Santa Fe Institute, 1399 Hyde Park Road, Santa Fe, NM 87501, USA}
\date{\today}

\maketitle

\begin{center}
``Neither a borrower nor a lender be''\\
{\it Hamlet,} Act 1, Scene 3, 75
\end{center}

\begin{abstract}
\citet{Peters2011a} defined an optimal leverage which maximizes the time-average growth rate of an investment held at constant leverage. It was hypothesized that this optimal leverage is attracted to 1, such that, \eg, leveraging an investment in the market portfolio cannot yield long-term outperformance. This places a strong constraint on the stochastic properties of prices of traded assets, which we call ``leverage efficiency.'' Market conditions that deviate from leverage efficiency are unstable and may create leverage-driven bubbles. Here we expand on the hypothesis and its implications. These include a theory of noise that explains how systemic stability rules out smooth price changes at any pricing frequency; a resolution of the so-called equity premium puzzle; a protocol for central bank interest rate setting to avoid leverage-driven price instabilities; and a method for detecting fraudulent investment schemes by exploiting differences between the stochastic properties of their prices and those of legitimately-traded assets. To submit the hypothesis to a rigorous test we choose price data from different assets: the \SP index, Bitcoin, Berkshire Hathaway Inc., and Bernard L. Madoff Investment Securities LLC. Analysis of these data supports the hypothesis.
\end{abstract}



\section{Introduction}
In section~\ref{Mathematical} we summarize a few key properties of
geometric Brownian motion that were pointed out in \citep{Peters2011a}.
We indicate the main elements of the analogy that is often drawn
between this model and the dynamics of markets. Section~\ref{Stochastic}
introduces the concept of leverage efficiency, namely the hypothesis
that the properties of price fluctuations in real markets are strongly
constrained by efficiency arguments so as to make
investments of leverage 1 optimal. This hypothesis is motivated by the
considerations in section~\ref{Mathematical} but goes beyond the simple
model discussed there. The arguments are robust enough to yield
insights into other assets, such as houses, or indeed national
economies or the global economy.
We summarize example applications of this fundamental constraint on price dynamics in section~\ref{Applications}: an explanation of asset price changes in the absence of new information; a resolution of the equity premium puzzle; a protocol for central bank rate setting to avoid leverage bubbles; and the detection of fraudulent investment schemes.
Section~\ref{Tests} confronts our theoretical work
with data. The arguments leading to the hypothesis
are neither specific to the mathematical model nor to any particular asset.  
Section~\ref{Discussion} concludes that optimal leverage is empirically close to 1, as predicted by leverage efficiency, for investments in assets as different as  the \SP index (\SPtick, 1927--2020) and Bitcoin (BTC, 2010--2020).

\section{Mathematical background}\label{Mathematical}
{\bf Notation:} The present study uses three different levels of
realism. To avoid tedious nomenclature and confusion between these, we
use three different superscripts:
\begin{enumerate}
\item
superscript $\supm$ refers to the mathematical model
used to motivate and guide our investigations;
\item 
superscript $\supc$ refers to data analyses performed to
test our main hypothesis empirically; and
\item
superscript $\supr$ refers to corresponding quantities in the
context of real markets and their participants.
\end{enumerate}

{\bf Model:}
The observable $x(t)$ is said to undergo geometric Brownian motion if it obeys the \Ito\ stochastic differential equation,
\begin{equation}
dx(t)=x(t)(\mum dt + \sigmam dW),
\elabel{motion}
\end{equation}
where $t$ denotes time, $dt$ its infinitesimal increment, and $dW$ the normally distributed Wiener increment, $dW\sim \mathcal{N}(0,dt)$. We call $\mum$ the drift and $\sigmam$ the volatility.

We pause here to make an important distinction between
averaging over an ensemble and averaging over time. For some special observables, 
this distinction is unimportant because they have the following 
ergodic property \citep{PetersGell-Mann2016,Peters2019b}:\\
\\
{\it \underline{Equality of averages}\\
The expectation value of the observable is a constant (independent of time) and the finite-time average of the observable converges to this constant with probability one as the averaging time tends to infinity.}\\
\\
The observable $x(t)$ defined by \eref{motion} does not possess this property. Therefore, we cannot
assume that the expectation value $\ave{x(t)}$ will be informative of what happens to $x(t)$ over time. From now on we will refer to the expectation value as the ``ensemble average'' because it has little to do with the everyday meaning of the word ``expect,'' whereas it is by definition the average over an ensemble of systems.


Consider the growth rate estimator,
\begin{equation}
\gestm(\Dt,N)\equiv\frac{1}{\Dt}\ln \left(\frac{1}{N}\sum_i^N\left(\frac{x_i(t+\Dt)}{x_i(t)}\right)\right),
\elabel{estimator}
\end{equation}
where $i$ indexes realizations of the process described by
\eref{motion} and $\Dt$ is a time increment. Taking the limit $N\to\infty$ for finite $\Dt$ extracts
the behavior of the ensemble average, whereas taking the limit $\Dt\to\infty$ for finite
$N$ extracts the long-time behavior \citep{PetersKlein2013,PetersAdamou2018b}. This procedure
yields clear interpretations of two well-known characteristics of
geometric Brownian motion:
\begin{equation}
\gens\equiv\lim_{N\to\infty}\gestm(\Dt,N)=\mum
\elabel{ensemble_growth}
\end{equation}
shows that the ensemble-average growth rate is equal to the drift; and
\begin{equation}
\gtm\equiv\lim_{\Dt\to\infty}\gestm(\Dt,N) = \ave{\frac{\Delta\ln x}{\Dt}} = \mum-\frac{({\sigmam})^2}{2} 
\elabel{time_growth}
\end{equation}
shows that the long-time growth rate, \ie what an individual will experience eventually, is smaller by a correction term $(\sigmam)^2/2$. \Eref{time_growth}, which shows $\gtm$ as equal to the ensemble average of the rate of change $\Delta\ln x / \Delta t$, is obtained by applying \Ito's formula to \eref{motion}. Indeed, this rate of change is an ergodic observable for the multiplicative dynamics defined by \eref{motion}, as discussed in \citep{PetersGell-Mann2016}.


Referring to the analogy with stock markets, it was pointed out in \citep{Peters2011a} that an individual investor should be more concerned about $\gtm$, the long-time growth rate of a single realization of the process, than about $\gens$, the growth rate of the average over parallel realizations which are inaccessible to him. For historical reasons, however, $\gens$ is often mistakenly considered in the literature, as noted in \citep{HughsonETAL2006}.

We introduce a leverage parameter $\lm$ to extend the market analogy to leveraged investments. We imagine two assets available to an investor: one riskless, whose price $x_\text{riskless}(t)$ obeys \eref{motion} with drift $\murm$ and volatility zero; and the other risky, whose price $x_\text{risky}(t) $ obeys \eref{motion} with drift
\be
\mum=\murm+\muem
\ee
and volatility $\sigmam>0$. The investor has net resources of $\xl(t)$ to allocate between these assets. A leveraged investment in the risky asset is, in effect, a portfolio in which $\lm \xl$ is held in the risky asset and the remainder $(1-\lm)\xl$ is held in the riskless asset. Each holding achieves the same fractional change -- which we shall call the ``return'' -- as its respective asset, so that the total resources evolve as
\bea
d\xl &=& (1-\lm) \xl \left(\frac{dx_\text{riskless}}{x_\text{riskless}}\right) + \lm \xl \left(\frac{dx_\text{risky}}{x_\text{risky}}\right) \\
&=& \xl [(\murm+\lm \muem) dt + \lm \sigmam dW ].
\elabel{l_motion}
\eea
The case $\lm=0$, in which the investor holds only the riskless asset, results in deterministic exponential growth at a rate equal to the drift $\murm$. The case $\lm=1$, in which the investor places all his resources in the risky asset, is equivalent to \eref{motion}, where $\mum=\murm+\muem$.

We note that $\lm$ is constant in this setup. The fractional holdings in the two assets do not change over time, even though the values of the holdings do change. Unless $\lm=0$ or $1$, when only one asset is present, this implies that the portfolio is continuously rebalanced, \ie resources are moved between the risky and riskless assets to maintain the leverage at $\lm$. This rebalancing trade is shown schematically in \fref{rebalance}. In practice, such rebalancing would take place only at finite time intervals and would incur transaction costs. Such effects are included in our empirical work in section~\ref{Tests}. Moreover, the leveraged investments imagined in this study should not be confused with ``buy-and-hold'' portfolios which start with an initial leverage and do not undergo rebalancing. In general, the allocation of such portfolios will change over time.
\begin{figure}
\centering
\includegraphics[width=.8\textwidth]{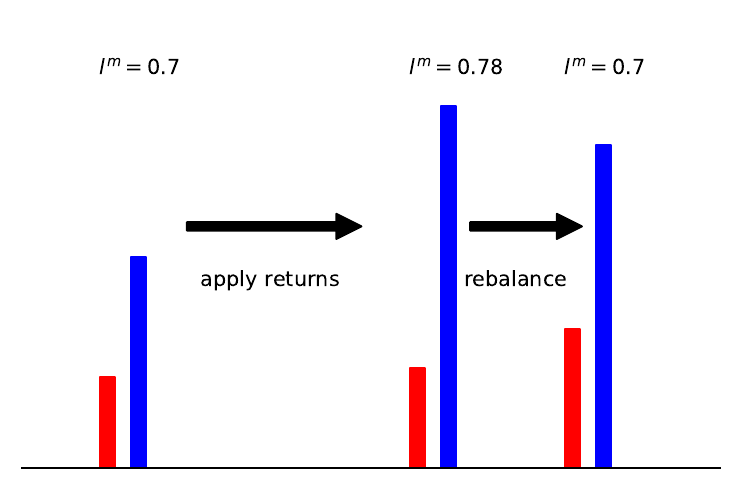}
\caption{Rebalancing of a portfolio containing risky (blue) and riskless (red) assets. Leverage starts at $\lm=0.7$, meaning 70\% of the portfolio value is invested in the risky asset and 30\% in the riskless asset. Each investment experiences the relative returns of the corresponding asset. In this case, the risky asset does better than the riskless asset, causing the leverage to increase temporarily to $0.78$. To return the leverage to $0.7$, some risky asset must be traded for riskless asset in a so-called rebalancing trade.
\flabel{rebalance}}
\end{figure}

For the moment, we will think of the risky asset as resembling an investment in the market portfolio and of the riskless asset as resembling a safe government bond or bank deposit. Negative holdings, corresponding to borrowed assets, are possible. Leverage $\lm<0$ reflects short-selling; $0\leq\lm\leq1$ reflects part of the investor's equity being invested in the market
and part kept safe accruing interest at rate $\murm$; and $\lm>1$ reflects
what is commonly referred to as leveraging, \ie an investment in the
market that exceeds the investor's equity and includes borrowed
funds. The volatility in \eref{l_motion} is $\lm\sigmam$, proportional to the leverage, and the drift is $\murm+\lm\muem$, reflecting a safe interest
rate and the excess drift of the market added in proportion
to the leverage. Thus leveraging causes both the excess drift and the
fluctuation amplitude to increase linearly.

 $\xl(t)$ in \eref{l_motion} has the leverage-dependent ensemble-average growth rate,
\begin{equation}
\gens(\lm)=\murm+\lm\muem,
\elabel{l_ensemble_growth}
\end{equation}
and the leverage-dependent time-average growth rate,
\begin{equation}
\gtm(\lm)=\murm+\lm\muem-\frac{(\lm\sigmam)^2}{2}.
\elabel{l_time_growth}
\end{equation}
Crucially \eref{l_time_growth}, unlike \eref{l_ensemble_growth}, is not monotonic in $\lm$. Maximizing $\gtm(\lm)$ establishes the existence of an objectively optimal leverage:
\begin{equation}
\elabel{lopt}
\loptm=\frac{\muem}{(\sigmam)^2}.
\end{equation}

\Eref{lopt} implies that, unless $\muem/(\sigmam)^2=1$, it is
possible to choose $\lm$ in \eref{l_motion} such that $\xl(t)$
(reflecting a leveraged investment)
outgrows $x_\text{risky}(t)$ in \eref{motion} (reflecting the market
portfolio) in the long run. For example, $0<\loptm<1$ would imply that, due to the nonlinear effects of multiplicative fluctuations, a rising market could be beaten by keeping a fixed fraction of one's resources in a savings account.

In reality, the outcome of an investment held for some finite time $\Dt$ is given by the growth of the investment averaged over $\Dt$. The growth rate of the ensemble average is {\it a priori} irrelevant in practice. 
Maximizing $\gens$ in \eref{l_ensemble_growth} leads to the recommendation of maximizing $\lm$ (or $-\lm$).
Absent constraints on leverage, this would lead to a negative divergence in $\gtm$, \ie to ruin, in \eref{l_time_growth}. Thus, if $\gens$ (often called the ``expected rate of return'') is falsely believed to reflect the quantity an investor should
optimize, and if $\lm$ is interpreted as the leverage used in the
investment, then the investor would be led to exceed (positively or
negatively) the leverage that would truly be most beneficial, with potentially deleterious effects on wealth.

The history of the struggle to make sense of the misleading recommendations derived from \eref{l_ensemble_growth} is the history of decision theory and of probability theory itself,
see \citep{Peters2011b,PetersGell-Mann2016,Peters2019b}. Under multiplicative growth, such as in \eref{motion}, the difference between the growth rate of the ensemble average and the long-term growth rate of an individual trajectory is the difference between arithmetic and geometric means. This was identified in the context of a repeated gamble as early as \citep{Whitworth1870}, while the correction term in its present form, $-(\sigmam)^2/2$, follows directly from \citep{Ito1944}. Optimal leverage for \eref{l_motion} was computed by \citep{Kelly1956} and in the form of \eref{lopt} by \citep{Merton1969}, although neither pointed to the non-ergodicity of $x(t)$ as the origin of their findings. The present study is concerned with the dynamic properties of the optimal leverage observed in time series from real markets.

\section{Leverage efficiency}\label{Stochastic}
The efficient market hypothesis \citep{Bachelier1900b,Fama1965} claims that the price of an asset traded in an efficient market reflects all the information publicly available about the asset. The corollary is that it is impossible for a market participant, without access to privileged information, consistently to achieve growth at a rate exceeding the long-time growth rate of the market (``to beat the market'') by trading assets. We refer to this concept as {\it price efficiency}.

Here we propose a different, fluctuations-based, market efficiency, which we call\\
\\
{\it \underline{Leverage efficiency:}\\
It is impossible for a market participant without privileged information to beat the market
by applying leverage.}\\
\\
Simple strategies such as borrowing money to invest, $\lr>1$, or keeping some money in the bank, $\lr<1$, should not yield consistent market outperformance, \ie there should be no leverage arbitrage. This reasoning was used in \citep{Peters2011a} to hypothesize that real markets self-organize so that
\begin{equation}
\loptr=1
\elabel{hypothesis}
\end{equation}
is an attractive point for their stochastic properties (represented by
$\murm$, $\muem$, and $\sigmam$ in the model).

The hypothesis we are about to test is motivated by the model
\eref{l_motion} and its properties in \eref{l_ensemble_growth},
\eref{l_time_growth}, and \eref{lopt}, in the sense that this model
motivates the existence of an optimal leverage. But it is by no means
derived from the model, as the hypothesis requires the dynamic
adjustment, or self-organization, of the stochastic properties of the
system, which, in the model, are represented by fixed parameters. One
would have to think of $\murm, \muem$, and $\sigmam$ as slowly-varying (compared
to the fluctuations) functions of time, related to one another as well as to $\lm$ through a dynamic which has $\loptm=1$, \ie
\begin{equation}
\frac{\muem}{(\sigmam)^2}=1,
\elabel{hypothesis_m}
\end{equation}
as an attractor.

Although inspired by a mathematical model, the hypothesis in \eref{hypothesis} does not rest on model-specific properties. Crucial for it are the identification of the time-average growth rate 
and the consequent establishment of an optimal leverage, about which economic arguments may be framed.

Leverage efficiency is a tantalizing concept. It posits that the market has a different quality of knowledge than implied by price efficiency. Price efficiency is essentially a static concept, which states that prices coincide with some notion of objective value. Leverage efficiency, on the other hand, constrains dynamics and predicts properties of fluctuations of the prices of traded assets.

Leverage efficiency arises from a dynamical feedback in which prices and their fluctuations respond to changes in optimal leverage, in a manner reminiscent of the basic feedback between prices and supply-demand imbalances familiar in economics. We augment this with criteria for global stability derived from the no leverage arbitrage argument. This mechanism, detailed below, suggests that both $\loptr=1$ and $\loptr=0$ are particularly attractive and that the interval $0\leq\loptr\leq 1$ constitutes a stable regime, whereas values outside it are unstable.

\begin{enumerate}
\item
\textbf{Leverage feedbacks:} 
    \begin{enumerate}
    \item
If $\loptr>1$, investors will eventually borrow money to invest. High demand for risky assets will lead to price increases and low demand for riskless deposits will lead to increased yields on safe bonds, $\murr$. Both effects reduce $\muer$. In addition, highly leveraged investments are liable to margin calls and tend to increase volatility. The fall in $\muer$ and the rise in $\sigmar$ act to decrease $\loptr$.
    \item
If $0<\loptr<1$, there is no feedback. If we imagine leverage decreasing from scenario (a), asset prices fall and bond yields drop as investors withdraw from the market and move resources to safe deposits. Both effects increase $\muer$. Optimally leveraged investments require no borrowing, so volatility-increasing margin calls do not occur. Thus the pressures in (a) bearing down on $\loptr$ are relaxed. This regime is marginally stable.
    \item
    If $\loptr<0$, investors will eventually borrow stock to
    short-sell. Low demand for risky assets will lead to price decreases and high demand for riskless deposits to a decrease in yields on safe bonds $\murr$.
    Both effects make $\muer$ less negative. Highly negatively leveraged investments are liable to
    margin calls and tend to increase volatility. The 
    increase in $\muer$ and the rise in $\sigmar$ imply that $\loptr$ becomes less negative.
    \end{enumerate}
\item
\textbf{Global stability:} It is difficult to envisage globally stable
economies existing with optimal leverage outside the interval
\mbox{$0\leq\loptr\leq 1$} because:
    \begin{enumerate}
    \item
    If $\loptr>1$, everyone should invest in the market more than he
    owns. This is not possible because the funds to be invested must
    be provided by someone. 
    \item
    If $\loptr<0$, everyone should sell more market shares than he
    owns. This is not possible because the assets to be sold must be
    provided by someone. 
    \end{enumerate}
Thus the range $0\leq\loptr\leq 1$ is special in not being globally
unstable.
\end{enumerate}
We believe the above to be the main drivers behind leverage
efficiency. There are additional effects, however, which
reinforce it.
\begin{enumerate}
\item
\textbf{Economic paralysis:} In an economy with $\loptr\leq 0$ there is
no incentive to invest in risky assets, which may limit productive economic
activity. Policy makers will tend to steer away from such conditions, perceiving 
$\loptr=1$ as more desirable.
\item
\textbf{Covered short-selling:} An investment with $\lr<0$ is punished
by the costs of borrowing stock to short-sell, \ie covered as opposed
to naked short-selling.
\item
\textbf{Asymmetric interest rates:} The interest received by a depositor is
typically less than the interest paid by a borrower. Therefore, an
investment with $\lr<1$ is punished by low deposit interest rates and
an investment with $\lr>1$ is punished by high borrowing costs. This
reinforces $\loptr=1$ as an attractive point.
\item
\textbf{Transaction costs:} The costs of buying and selling assets
(fees, spreads, and so on) punish any strategy that requires
trading. Holding an investment of constant leverage generally requires
trading to rebalance the ratio of assets to equity. The two
exceptions are investments with $\lr=0$ and $\lr=1$.
\end{enumerate}


\subsection{Predictive accuracy}
If we use only observed price changes to compute changes in invested wealth, we will tend to overestimate the magnitude of optimal leverage, \ie $|\loptc|>|\loptr|$. This is because real leveraged investments incur trading and borrowing costs. Therefore, if the leverage efficiency hypothesis holds, we would expect a na\"ively determined value of $\loptc$ to be a little larger than its predicted value of $\loptr=1$. It is difficult to estimate the size of this bias, although we will attempt to do so in Sec.~\ref{Tests}.

Ignoring these effects for the moment, we can predict how close we expect observed optimal leverage, $\loptc$, to be to its predicted value of 1, when we estimate it from a finite time series. Even assuming that our theory is correct and 
$\loptr$ is attracted to a particular value, we expect random
deviations from it to increase as the time series gets shorter. To take an extreme example, with daily data and assuming daily rebalancing, the observed optimal leverage over a
window of one day does not exist. Either $\loptc \to+\infty$ if the
risky return beats the riskless return on that day; or $\loptc \to -\infty$ if
riskless beats risky. Indeed, the magnitude of the observed
optimal leverage will diverge for any window over which the daily
risky returns are either all greater than, or all less than, the daily riskless returns. This is unlikely for windows of months or years but using daily data it happens
commonly over windows of days or weeks. Even without this
divergence, shorter windows are more likely to result in larger
positive and negative optimal leverages because relative fluctuations
are larger over shorter time scales.

To quantify this idea  we solve \eref{l_motion} for the time-average growth rate after a finite
time $\Dt$. The noise term, which vanishes in the limit $\Dt\to\infty$, is retained to give
\begin{equation}
\widehat{\glm}(\Dt,N=1) =
\murm+\lm\muem-\frac{(\lm\sigmam)^2}{2}+\frac{\lm\sigmam W(\Dt)}{\Dt}.
\end{equation}
Maximizing this generates a noisy estimate for the optimal leverage over a window of length $\Dt$,
\begin{equation}
\loptestm(\Dt, N=1) = \loptm + \frac{W(\Dt)}{\sigmam \Dt}.
\end{equation}
Thus, in the model, the optimal leverage for a finite time series is
normally distributed with mean $\loptm$ and standard deviation
\begin{equation}
\text{stdev}(\loptestm(\Dt, N=1)) = \frac{1}{\sigmam \sqrt{\Dt}}.
\elabel{lopt_sd}
\end{equation}
We will use this quantity as the standard error for the prediction $\loptr=1$.

\section{Applications of leverage efficiency}\label{Applications}
\subsection{A theory of noise}\label{Noise}
Leverage efficiency sheds light on the origin and nature of ``noise'' in financial markets. According to leverage efficiency, prices of risky assets must fluctuate if an excess drift exists, $\muer>0$, simply because the market would otherwise become unstable. Furthermore, leverage efficiency tells us, in \eref{noise} below, the fluctuation amplitude required to avoid instability. But, if price fluctuations are necessary for stability, then the intellectual basis for price efficiency -- that changes in price are driven by the arrival of new economic information --  cannot be the whole truth. At least some component of observed fluctuations must be driven by the leverage feedbacks described in section~\ref{Stochastic}, which enforce leverage efficiency and which have little to do with information arrival.

\citet{Black1986} differentiated between information-based and other types of price fluctuation, referring to the latter as ``noise'' and regarding it as a symptom of inaccurate information and market inefficiency. In our model, rearranging \eref{hypothesis_m} yields a fluctuation amplitude of
\be
\elabel{noise}
\sigmam=\sqrt{\muem}.
\ee
Thus, based purely on leverage efficiency, prices must fluctuate and we can even quantify by how much. An asset whose expectation value grows faster than that of the riskless asset must fluctuate, otherwise systemic stability will be undermined.
This is a radical departure from conventional thinking, which has practical consequences. For instance, prices ``discovered'' at ever higher trading frequencies must necessarily reveal more ups and downs, but this noise is self-generated, imposed by the requirement of leverage stability. Stability as the genesis of volatility constitutes a theory of noise requiring no appeal to the arrival of unspecified information, whether accurate or not.

\subsection{Equity premium puzzle}\label{EPP}
The term ``equity premium'' describes a form of compensation investors demand for holding a risky asset. One way of quantifying this idea is as follows. Imagine you hold a riskless asset with a given time-average growth rate, equal to its drift, which you may swap for a risky asset with a given volatility and a larger drift. How much greater must the risky asset's time-average growth rate be to entice you to swap? 

The literature on this question is large and often takes a psychological and individual-specific perspective. For instance, a more ``risk averse'' individual will demand a higher equity premium to hold the risky asset. Models of human behaviour enter both into the definition of the equity premium -- which, as noted by \citet{Fernandez2009}, lacks consensus -- and into its analysis. Much of the literature draws the conclusion that observed equity premia are inconsistent with dominant behavioural models, typically being larger than those models predict. This is known as the ``equity premium puzzle'' \citep{MehraPrescott1985}. \cite{LeRoy2016} summarizes the current state of the debate as follows: ``Most analysts believe that no single convincing explanation has been provided for the volatility of equity prices. The conclusion that appears to follow from the equity premium and price volatility puzzles is that, for whatever reason, prices of financial assets do not behave as the theory of consumption-based asset pricing predicts.''\footnote{The ``price volatility puzzle'' is that prices are apparently more volatile than dominant models predict.}

The framework we have developed here takes a psychologically na\"ive perspective. We define the equity premium, $\pim$, without reference to human behaviour or consumption models. It is the difference between the time-average growth rates of the risky asset ($\lm=1$) and the riskless asset ($\lm=0$), which in our model is
\bea
\pim &\equiv& \gtm(1) - \gtm(0)\\
&=& \muem-\frac{(\sigmam)^2}{2}.
\elabel{epdef}
\eea
We ask what value we expect the equity premium to take in a real market. Leverage efficiency dictates how large volatility must be for the market to avoid a leverage instability (and, therefore, to survive so that we can ask the question). Substituting \eref{hypothesis_m} in \eref{epdef} reveals that, under leverage efficiency, the real equity premium is attracted to
\be
\pim = \frac{(\sigmam)^2}{2}.
\elabel{epval2}
\ee
or, equivalently, to
\be
\pir = \frac{\muer}{2}.
\elabel{epval1}
\ee
Our data analysis in section~\ref{Tests} confirms this prediction.

Under realistic assumptions about actual trading costs, observed optimal leverage is very close to one, meaning that volatility has just the right magnitude to prevent leverage-driven bubbles without making investments in the risky assets unattractive. If trading costs are ignored, $\loptc$ is greater than one, meaning volatility is a little lower than na\"ively predicted.
We regard the consistency of the observed equity premium with the leverage efficiency hypothesis as the long-sought resolution of the equity premium and price volatility puzzles.

\subsection{Central bank rate setting}\label{Rate_setting}
Our observations are also relevant to a central bank setting its lending rate. The rate setter would view the total drift $\mur$ of an appropriate asset or index as given, and the risk-free drift $\murr$ as the central bank's rate. If the aim is to achieve full investment in productive activity without fuelling an asset bubble, then this rate should be set so that $\loptr=1$. Since $\loptr=\muer/(\sigmar)^2$ and $\muer=\mur-\murr$, this is achieved by setting 
\be
\murr=\mur-(\sigmar)^2.
\ee
Using $\muc$ and $\sigmac$ for the \SP index in section~\ref{TheEntire}, the optimal interest rate comes out as 3.8\% {\it p.a.} for the period analyzed (1927--2020). This is similar to typical real interest rates over this period, suggesting that rate-setting may have been informed, at least in part, by the considerations we outlined above. The task of the central banker can be seen as the task of estimating $\mur$ and $\sigmar$ in the relevant way. This will involve choices about data and timescales which are far from trivial. For instance, in our data analyses at any given time there is an estimate for $\muc$ and one for $\sigmac$ for each possible length of lookback window. Operational matters aside, stability with respect to leverage is an important consideration for any central bank. Leverage efficiency provides a simple quantitative basis for a rate setting protocol and may frame qualitative discussions about interest rates in a useful way.

\subsection{Fraud detection}\label{Fraud}
The arguments advanced in support of the leverage efficiency hypothesis in section~\ref{Stochastic} do not relate to specific pairs of riskless and risky assets. Rather, they are general arguments about assets traded in markets, and we expect the predictions of leverage efficiency to hold generally for such assets, absent special conditions. Specifically, we expect observed optimal leverage, $\loptc$, over finite time, $\Dt$, to be consistent with the predictions of its value and uncertainty in \eref{hypothesis} and \eref{lopt_sd}, \ie
\be
\loptr = 1 \pm \frac{1}{\sigmar \sqrt{\Dt}},
\elabel{lopt_range}
\ee
where this range corresponds to one standard deviation in each direction in the model.

Observed optimal leverages inconsistent with this prediction suggest two possibilities: either the prediction is wrong generally; or there are special circumstances associated with the assets under study which make it wrong specifically. We offer evidence in section~\ref{Tests} to refute the first possibility, so we turn our attention to the second. In what circumstances would we expect to observe anomalously high or low optimal leverages? Perhaps when we pick an asset with atypically good or bad performance over the time period studied (which, indeed, we do deliberately using Berkshire Hathaway in section~\ref{Tests}). Or perhaps when we study an asset class for which economic policies, such as lower or higher taxes, exist to promote or discourage investment.

Another circumstance, our focus here, is a fraudulent asset whose reported prices are not the result of the trading feedbacks described in section~\ref{Stochastic}. We do not expect our predictions to hold for fabricated asset prices (unless the fabricator is aware of this work). This suggests that we might use observed deviations from leverage efficiency, \ie inconsistencies with \eref{lopt_range}, as a method for detecting fraudulent investments. We test this proposal in section~\ref{Tests} on the infamous Madoff Ponzi scheme.

\section{Tests of leverage efficiency in historical data}\label{Tests}
We test the leverage efficiency hypothesis by computing the growth rates of constant-leverage investments in the \SP index, Bitcoin, Berkshire Hathaway (class A shares), and Madoff's Ponzi scheme. We assume that cash attracts interest at the Federal Reserve overnight rate.

\subsection{Data sets and codes}
The data used in this study are publicly available.
\begin{itemize}
\item \SP (\SPtick): from \href{https://finance.yahoo.com/quote/%5ESPX/history?p=%5ESPX}{https://finance.yahoo.com/}
\item Bitcoin (BTC): from \href{http://www.coindesk.com/price/}{http://www.coindesk.com/price/}
\item Berkshire Hathaway (BRK):  from \href{https://uk.finance.yahoo.com/quote/BRK-A}{https://finance.yahoo.com/}
\item Madoff (MAD):  from \href{http://law.du.edu/documents/corporate-governance/legislation/Markopolos-Madoff-Complaint.pdf}{http://law.du.edu/}, digitized at \href{http://lml.org.uk/wp-content/uploads/2019/12/madoff.csv}{http://lml.org.uk/madoff/}
\item Short-term federal rates (FED): spliced from \href{https://t.co/FDm5p3P828?amp=1}{} and \href{https://fred.stlouisfed.org/series/FEDFUNDS}{https://fred.stlouisfed.org}
\item \SP Total Return (SP500TR): from \href{https://finance.yahoo.com/quote/%5ESP500TR/history/}{https://finance.yahoo.com}
\item 10-year US Treasury bond rates (DGS10): from \href{https://fred.stlouisfed.org/series/DGS10}{https://fred.stlouisfed.org}
\end{itemize}

All of these data sets have their problems. Nonetheless they give an impression of how closely leverage efficiency manifests itself in real markets.

The \SP is an index of five hundred large companies, listed publicly in the United States. We use it as a proxy for a generic diversified investment in US stocks, but we note some caveats. Firstly, the index does not account for dividends paid to stockholders. This means it will tend to underestimate the performance of a real investment. Secondly, the index suffers from survivorship bias, representing a portfolio of the largest and most successful companies in the US, in which less successful companies are routinely replaced. This acts in the opposite direction to the first caveat. The available time series is sufficiently long (92 years) to leave reasonably small uncertainties in the prediction of optimal leverage, in which the standard error is 0.6.

The large volatility of BTC permits an even more precise prediction of optimal leverage, with a standard error of 0.3, despite the shortness of this time series (10 years). There is no consensus over what asset class BTC is, so it is remarkable that it behaves so similarly to more familiar assets.

Berkshire Hathaway, BRK, is a large conglomerate, well known for its sustained rapid growth over the last half century. We include it as an example of a successful and, we assume, legitimate business. As a cherry-picked investment, we anticipate that its optimal leverage will exceed 1, although by how much is an important question for our theory.

MAD -- Bernie Madoff's Ponzi scheme -- is very interesting. Here is a fictitious asset, whose returns were concocted to defraud investors. We find its behavior measurably different from properly traded assets, supporting our proposal in section~\ref{Fraud} that fraudulent investments may be detected by their deviations from the predictions of leverage efficiency.

The FED data are overnight interest rates paid between banks. It is unclear how well an investment in cash or bonds (or whatever one considers a ``riskless'' asset) is reflected by these rates. For instance, due to falling interest rates, a longer-term government bond would have appreciated considerably in recent decades. To the extent that short-term inter-bank rates are typically lower than real deposit and borrowing rates, we expect the FED data to underestimate the performance of riskless assets and, therefore, to overestimate the optimal allocation to risky assets.

The \SP total return includes dividends. Its time series is shorter, but we include it for completeness and to establish bounds on the effects of dividends.

The 10 year bond yields, DGS10, help us estimate a range of plausible optimal leverage values that might result from using different bond portfolios. Price changes of bonds are not taken into account in any of the analyses we present.

We suspect that, on balance, the competing biases in the data will tend to produce overestimates of real optimal leverage, $\loptc>\loptr$. 
We encourage readers to repeat the analyses for different assets and data sets, and to vary parameters and assumptions used in the data analysis. 
To facilitate this, we provide open-source Python code and data at \href{https://github.com/LMLhub/leverage_efficiency_codes}{https://github.com/LMLhub/leverage\_efficiency\_codes}.

\subsection{Data analysis}\label{Simulations}
The performance of an investment of constant leverage over a given time window is computed as follows. We start with equity of \$1, consisting of \$$\lc$ in the risky asset
and cash deposits of \$$(1-\lc)$. Each day the
values of these holdings are updated according to
historical riskless and risky returns. The
portfolio is then rebalanced, \ie the holdings in the risky asset are
adjusted so that their ratio to the total equity remains $\lc$. On
non-trading days the return of the market is zero, whereas deposits
continue to earn interest payments, which leads to an unrealistic
but negligible rebalancing on those days. The investment proceeds in
this fashion until the final day of the window, when the
final equity is recorded, \fref{trajectories}. 
\begin{SCfigure}
\centering
\includegraphics[width=.6\textwidth]{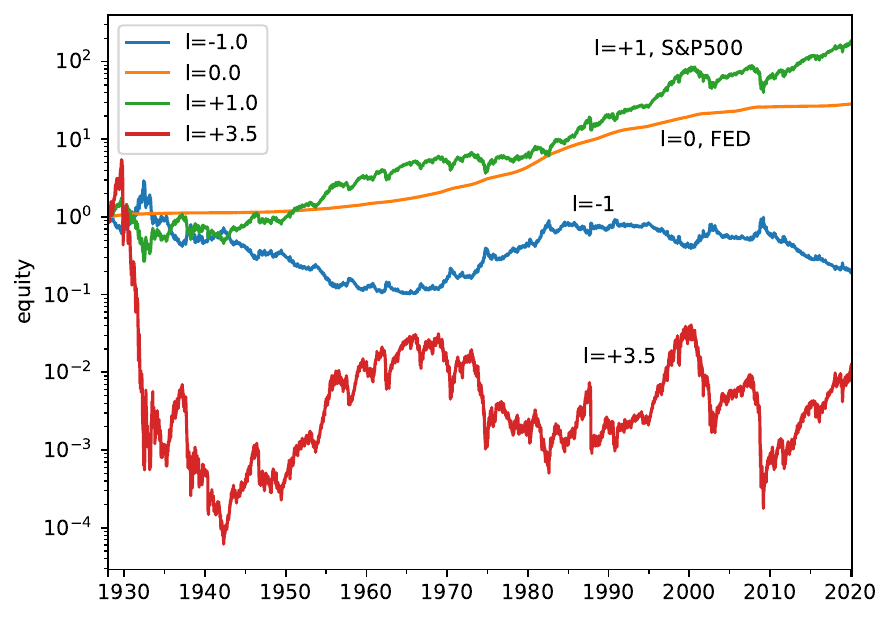}
\caption{Equity $\xlc(t; \lc)$ for investments of initially \$1 in the \SP at different constant leverages, where money is borrowed at overnight federal interest rates. As leverage increases, eventually the fluctuations become harmful and the investor loses money. For a given start date, each leverage produces one value for the final equity.
\flabel{trajectories}}
\end{SCfigure}

If at any time the total equity falls to or below
zero, we fix it there for all future times: the computation is considered bankrupt for the corresponding leverage,
\ie we do not allow recovery from negative equity. The optimal leverage, $\loptc$, is the leverage for which the final equity is maximized. 
\begin{figure}
\begin{picture}(300,130)(0,0)
\put(0,0){\includegraphics[height=.2\textheight]{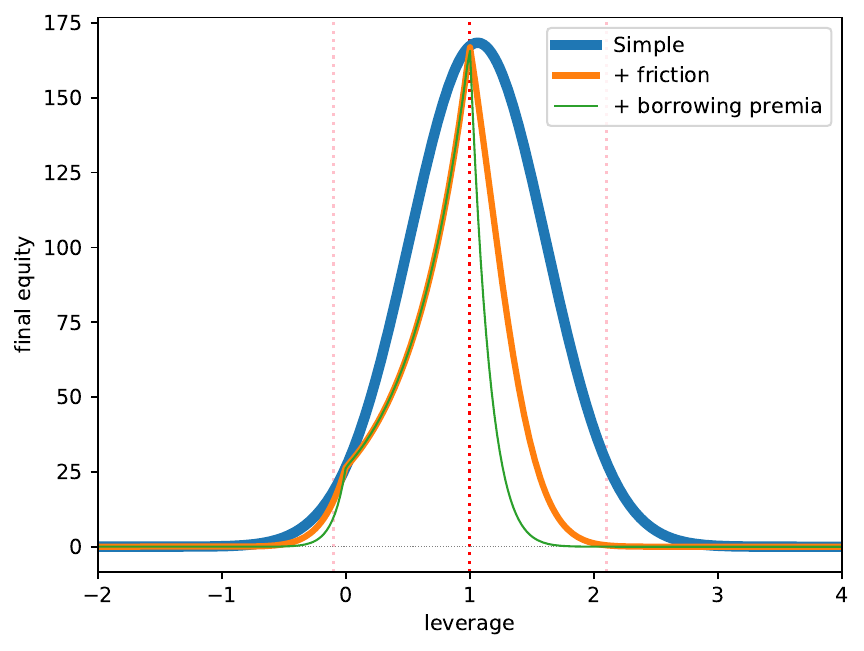}}
\put(180,0){\includegraphics[height=.2\textheight]{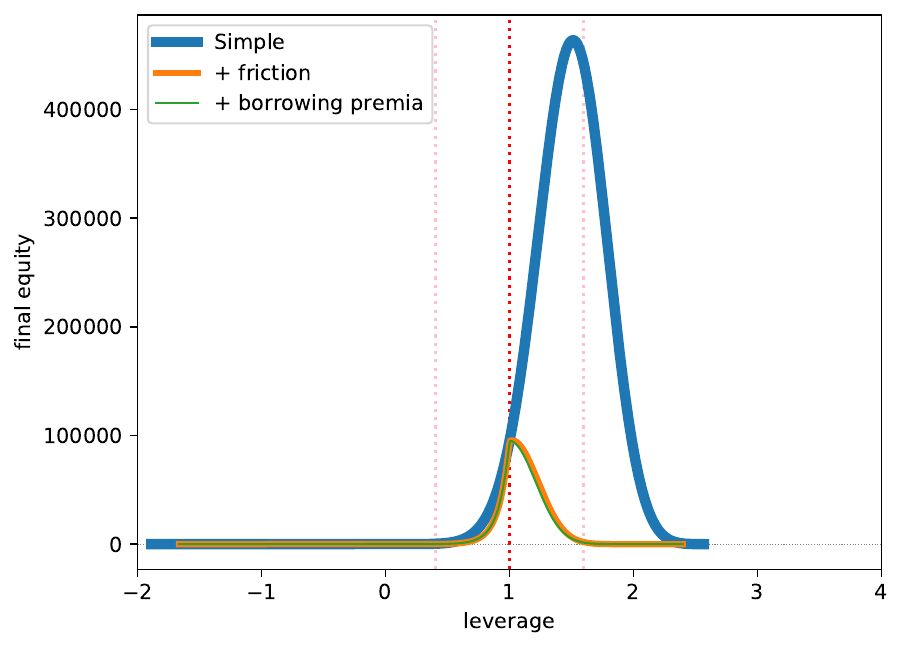}}
\put(40,130){a) \SP}
\put(230,130){b) Bitcoin}
\end{picture}
\caption{
Return-leverage curves for the \SP index and Bitcoin.\newline
Total return from constant-leverage investments in: a) \SPtick held from 1927-12-31 until 2020-03-01 (92 years); and b) BTC, held from 2010-07-19 and ending 2020-03-01 (10 years).
Data analyses 1 (blue), 2 (orange), 3 (green). For descriptions of the computations, see section~\ref{Simulations}.
Red vertical dotted lines indicate the prediction $\loptr=1$ and thin vertical pink dotted lines indicate two standard errors away from the prediction, see \eref{lopt_sd}.
The uncertainties in the prediction are larger for \SPtick because, while the time series is longer, the volatility is much lower than for BTC.
Bitcoin's volatility and drift are so high that \SPtick data going back more than 300 years would be required to achieve similar statistical accuracy.
}
\flabel{parabolic}
\end{figure}

\Fref{parabolic} shows the final equity (which has the same magnitude as the fractional change from start to finish) as a function of leverage for a hypothetical investment over the largest available window. The three curves in the figures correspond to the following three sets of assumptions about interest rates and transaction costs, mentioned as additional effects in section~\ref{Stochastic}:
\begin{itemize}
\item
{\bf Data analysis 1} (the ``simple'' case, blue line in \fref{parabolic}) is the least realistic analysis, in which the effective federal funds rate is applied to all cash,
whether deposited or borrowed.  No costs are incurred for
short-selling ($\lc<0$), akin to naked short-selling, \ie market
returns apply to negative stock holdings exactly as they apply
to positive holdings. Transaction costs are neglected. This results in
a smooth curve.
\item
{\bf Data analysis 2} (orange line) is like the
first case, but it includes friction. Whenever the portfolio is rebalanced a fraction of the value of the assets traded is lost to friction: 0.5\% for \SPtick, BRK, MAD; and 3\% for BTC. 
This resembles transaction costs and introduces kinks (\ie discontinuities in the first derivative of the curve) at $\lc=0$ and $\lc=1$.
\item
{\bf Data analysis 3} (the ``complex'' case, green line) is like the second case, but 5\% p.a.~risk premia are paid for borrowing cash to leverage up as well as for borrowing stock to go short.
The kinks become more pronounced.
\end{itemize}

\begin{figure}
\begin{picture}(300,130)(0,0)
\put(0,0){\includegraphics[height=.2\textheight]{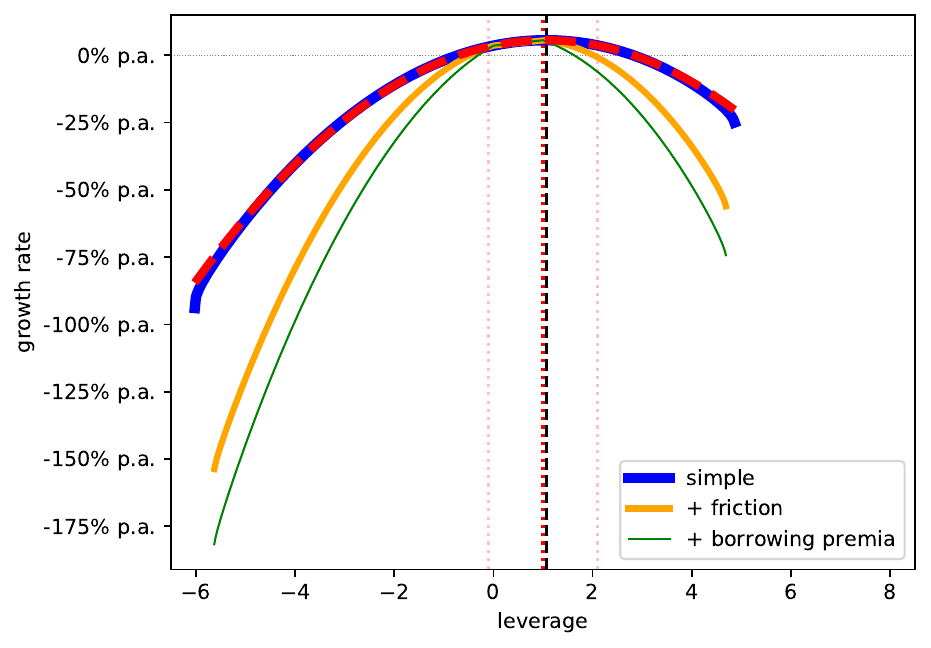}}
\put(180,0){\includegraphics[height=.2\textheight]{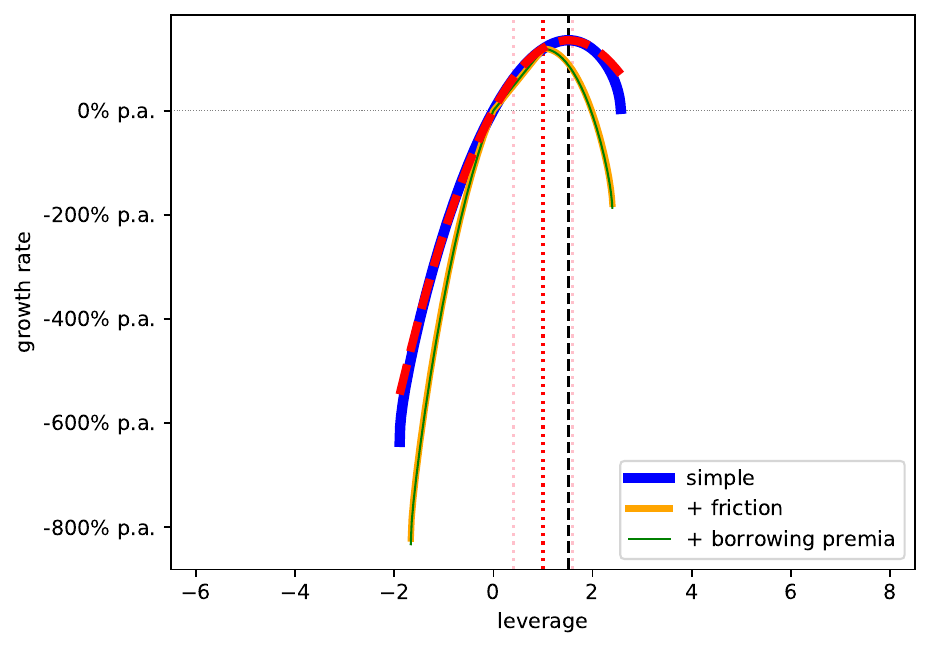}}
\put(80,130){a) \SP}
\put(260,130){b) Bitcoin}
\end{picture}
\caption{
Parabolic fits of leveraged time-average growth rates.\newline
Time-average growth rates as a function of leverage closely follow a parabola in backtests. a) \SPtick, b) BTC for the same time intervals as in \fref{parabolic}. Dashed red curves are parabolic fits on the central 80\% of the data. Deviations from the parabolic shape can be seen at the extremes, where the Gaussian model is deficient. Vertical dashed black lines indicate the empirical maximum; vertical dashed pink lines indicate the prediction for its position (optimal leverage 1) and two standard deviations out on either side.}
\flabel{parabolic2}
\end{figure}

As discussed in section~\ref{Stochastic}, covered short-selling, asymmetric interest rates, and transaction costs tend to penalize investments with leverages other than 0 or 1.
This is reflected in the empirical results by the kinks described above and visible in \fref{parabolic}.
For many time windows the discontinuity in the derivative of the return-leverage curve at $\lc=0$ or $1$ is accompanied by a change in sign of the derivative, making the point a global maximum and fixing $\loptc$ there.
The picture that emerges is this: based on price data only, but pretending that there are no costs to trading and borrowing, optimal leverage is greater than one.
For BTC, it is almost two standard errors greater, which we consider significant.
However, once our approximations of trading costs are included, real optimal leverage is seen to be very close or even equal to one. 

\subsection{The entire time series}
\label{TheEntire}
The return-leverage curves in \fref{parabolic} for an investment window spanning the entire time series shows optimal leverages of $\loptc=1.1$ (\SPtick) and $1.5$ (BTC) for the simple case (data analysis 1) and $\loptc=1.0$ (\SPtick) and $\loptc=1.0$ (BTC) for the complex case (data analysis 3). For comparison, predicted ranges for $\loptc$ were computed from \eref{lopt_sd} using the value of $\sigmac$ obtained by fitting \eref{l_time_growth} to return-leverage curves. These predictions were $\loptc=1\pm0.6$ (\SPtick) and $\loptc=1\pm0.3$ (BTC). The significance of these numbers will become clearer in section~\ref{ShorterTime}.

The time-average growth rate in \eref{l_time_growth}, specific to the model in \eref{l_motion}, is parabolic in $\lm$. We show in \fref{parabolic2} the time-average growth rate for the simple case, which is the logarithm of the total return divided by the window length. Given the known deficiencies of the geometric Brownian motion model, the parabolic fit (red dashed line) is remarkably good within the range of the model's validity. It is simultaneously remarkably bad outside this range: daily rebalanced investments with leverage $\lc<-6.0$ or $\lc>4.9$ (\SPtick) and $\lc<-1.9$ or $\lc>2.6$ (BTC) would have gone bankrupt (producing a negative divergence in the logarithmic return) due to extreme events. For highly-leveraged investors, the non-Gaussian tails of the return distribution, which determine their ruin probability, are far more important than any other property. 

The parameters of the fitted parabola are a fit of the model \eref{l_time_growth}. We take them as meaningful definitions of the empirical riskless drift $\murc$, excess drift $\muec$, and volatility $\sigmac$. We display them in \tref{leverage_numbers} for all the assets studied, together with the observed riskless growth rate $\gtc(0)$, observed equity premium $\pic$, and predicted equity premium $\muec/2$ from \eref{epval1}.

\begin{table}[h]
\begin{tabular}{c| c c c c c c c } 
& $\loptc$ & $\murc$ & $\muec$ & $\sigmac$ & $\gtc(0)$ & $\pic$ & $\muec/2$ \\ 
&  & (\%\pa) & (\%\pa) & (\%\psa) & (\%\pa) & (\%\pa) & (\%\pa)\\ 
\hline\hline
{\tiny 1927-12-31--2020-03-01} &&&&&&\\
SPX \vs FED & 1.1 (0.6)  & 3.5 &  3.8 & 19 & 3.6 & 2.0 & 1.9\\ 
{\tiny 2010-07-19--2020-03-01} &&&&&&\\
BTC \vs FED & 1.5 (0.3) & 1.2 & 180 & 110 & 0.66 & 120 & 89\\
{\tiny 1980-03-18--2020-03-01}  &&&&&&\\
BRK \vs FED  & 3.0 (0.7) & 4.3 & 16 & 23 & 4.5 & 13 & 7.9\\ 
{\tiny 1990-12-01--2005-05-01}  &&&&&&\\
MAD \vs FED & 100 (9) & -12 & 9.2 & 3.0  & 4.0 & 7.5 & 4.6\\ 
\hline
{\tiny 1962-01-02--2020-05-07}  &&&&&&\\
SPX \vs DGS10 & 0.6 (1.0) & 6.0 & 1.8 & 17  & 6.0 & 0.4 & 0.9\\ 

{\tiny 1988-01-05--2020-03-01}  &&&&&&\\
SP500TR \vs FED & 2.7 (1.0) & 3.0 & 8.3 & 18  & 3.2 & 6.4 & 4.1\\ 

\end{tabular}
\caption{{\bf Upper part:} key statistical properties of leveraged investments in SPX, BTC, BRK, MAD using FED rates. The number in brackets next to $\loptc$ are the standard error for the prediction. The first three rows show properly-traded assets. Optimal leverage, even without any adjustments for realistic trading costs, is within two standard errors of the predicted value of 1 for SPX and BTC. As anticipated, the estimates err on the high side. It is interesting that there is nothing remarkable about BTC in terms of optimal leverage: for systemic stability, an asset growing as fast as BTC must have volatility as large as that of BTC. BRK was selected for its impressive growth rate. Even an asset cherry-picked from the universe of publicly-traded assets, and without adjustment for friction, has an optimal leverage within three standard errors of the prediction. The fourth row -- the Madoff fraud -- represents an asset that never existed. Its observed optimal leverage is 100, fully eleven standard errors larger than the prediction of 1. This is caused by the lack of volatility in the reported returns, which should set off alarm bells. Values in this row, which depend on parabolic fits are meaningless because the monthly reporting of Madoff's returns makes proper rebalancing impossible. A better estimate for optimal leverage at higher rebalancing frequency is 200 to 300, see \fref{compare_assets}.} 
{\bf Lower part:} to gauge the effect of considering dividends and using overnight rates, we tried out SPX leveraged \vs ten-year rates, giving a low optimal leverage of 0.6; and SP500TR \vs overnight rates, giving a high optimal leverage of 2.7. 
\tlabel{leverage_numbers}
\end{table}

From a least-squares fit we obtain \mbox{$\murc=3.5\%$ \pa}, \mbox{$\muec=3.8\%$ \pa}, and \mbox{$\sigmac=19\%$ \psa} as the parameter estimates for \SPtick. For BTC the numbers are more interesting because they are outside the range of most people's intuition. Here we find \mbox{$\murc=1.2\%$ \pa} (interest rates were generally lower over the BTC period than the \SPtick period), \mbox{$\muec=180\%$ \pa}, and \mbox{$\sigmac=110\%$ \psa}. In both cases we performed a one-parameter fit on the central 80\% of the data in \fref{parabolic2}, after fixing the coordinates of the maximum.

The meanings of these numbers warrant some remarks. For \SPtick the riskless drift $\murc$ is practically identical to the realised time-average growth rate of a cash deposit. For BTC the estimate is a little off because the parabolic fit is not perfect and the scales of the process are so different that the riskless drift is irrelevant. To be explicit, the excess drift $\muec$ does not correspond to the excess growth rate of stock over cash, which we identify as the equity premium. Instead, due to the wealth-depleting effect of the volatility -- manifested in the model as $-(\lm \sigmam)^2/2$ in \eref{l_time_growth} -- a real investment in the \SP outgrew federal deposits at only \mbox{$\pic=2.0\%$ \pa}, which is less than the excess drift of \mbox{$\muec=3.8\%$ \pa}. For BTC the effect is vastly more pronounced: a real investment in BTC outgrew federal deposits by \mbox{$\pic=120\%$ \pa}.

We also include in \tref{leverage_numbers} statistical properties for two other assets, BRK and MAD, whose reported performance was notably impressive over the periods studied. BRK is an asset whose prices are the result of normal trading activity, while we know, thanks to \citet{Markopolos2005}, that MAD prices were fabricated to defraud investors. The observed optimal leverage of BRK is just under three standard errors greater than its predicted value of 1. This deviation from leverage efficiency, while significant, does not falsify our theory. We have deliberately chosen an asset with atypically rapid growth -- its performance is the stuff of investment legend -- to test the limits of our prediction.

By contrast, the observed optimal leverage of MAD is fully eleven standard errors greater than the prediction. This is a consequence of the extremely low volatility, \mbox{$\sigmac=3\%$ \psa}, reported by the fraudsters. These dramatic deviations from the prediction of leverage efficiency and from the volatility levels of other investment assets would, to a regulator aware of leverage efficiency, have indicated a need for further investigation. The difference between asset prices arising from trade and from fraud is further illustrated in \fref{compare_assets}: the growth rate-leverage curve for MAD is ``off the chart'' that would have been appropriate for \SPtick, BTC, and BRK.

\begin{SCfigure}
\centering
\includegraphics[width=.6\textwidth]{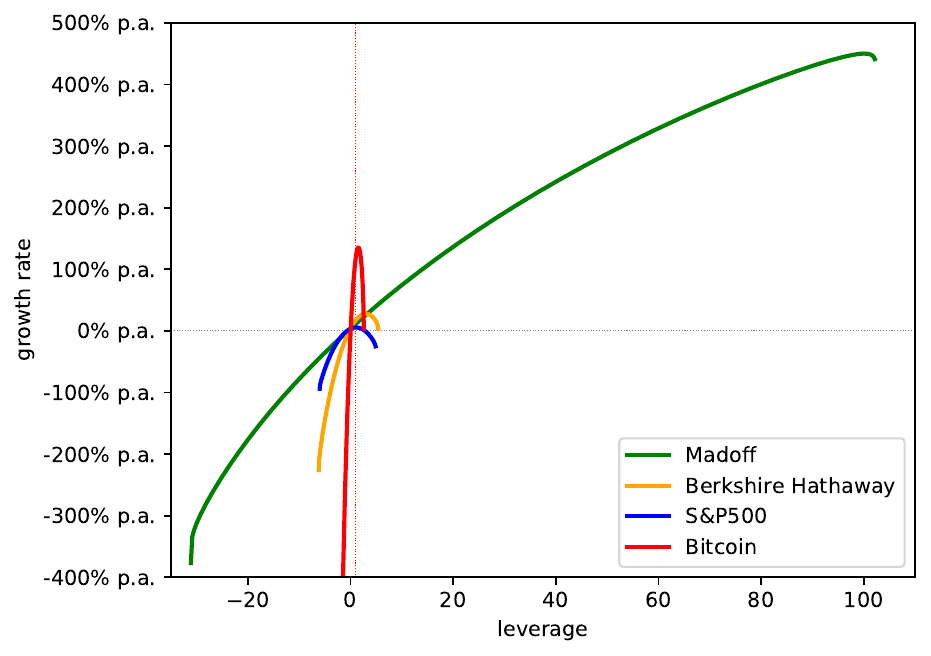}
\caption{
Parabolic fits of leveraged time-average growth rates for MAD (green), BRK (yellow), \SPtick (blue), and BTC (red).
The growth rate-leverage curve for the fraudulent asset, MAD, is on a wildly different scale to those for properly-traded assets.
\flabel{compare_assets}
}
\end{SCfigure}
\FloatBarrier

\subsection{Shorter time scales}
\label{ShorterTime}
So far we have used the predicted variance of observed optimal leverage, \eref{lopt_sd}, as a standard error. But we can go further by treating it as a testable prediction in itself.

Likening $\loptestm(\Dt, N=1)$ to the observed optimal leverage over a
window of size $\Dt$, we investigate how well \eref{lopt_sd} predicts the
fluctuations in $\loptc(\Dt)$. We compile histograms of $\loptc(\Dt)$ by
moving windows of size $\Dt$ across the record and compare the standard
deviation of $\loptc(\Dt)$ found in these histograms to the standard
deviation of $\loptestm(\Dt,N=1)$. 

\Fref{loglog} shows, on double-logarithmic scales, the standard
deviation of $\loptc$ against the window length for the simple
case. Good agreement is found with the model-specific prediction
in \eref{lopt_sd}. We note that, for shorter time scales, the
standard deviation is slightly higher than predicted, which may have to do 
with data discreteness and the divergence of optimal leverage for short windows with few rebalancing trades. 
For long windows the statistics are poor, primarily because we consider only non-overlapping windows.
There are no free parameters in \eref{lopt_sd}. The correct $\Dt^{-1/2}$-scaling is observed, and even the prefactor $1/\sigmac$ is approximately right ($\sigmac$ is the estimate from Sec.~\ref{TheEntire}).

\begin{figure}
\begin{picture}(300,250)(0,0)
\put(40,0){\includegraphics[width=.8\textwidth]{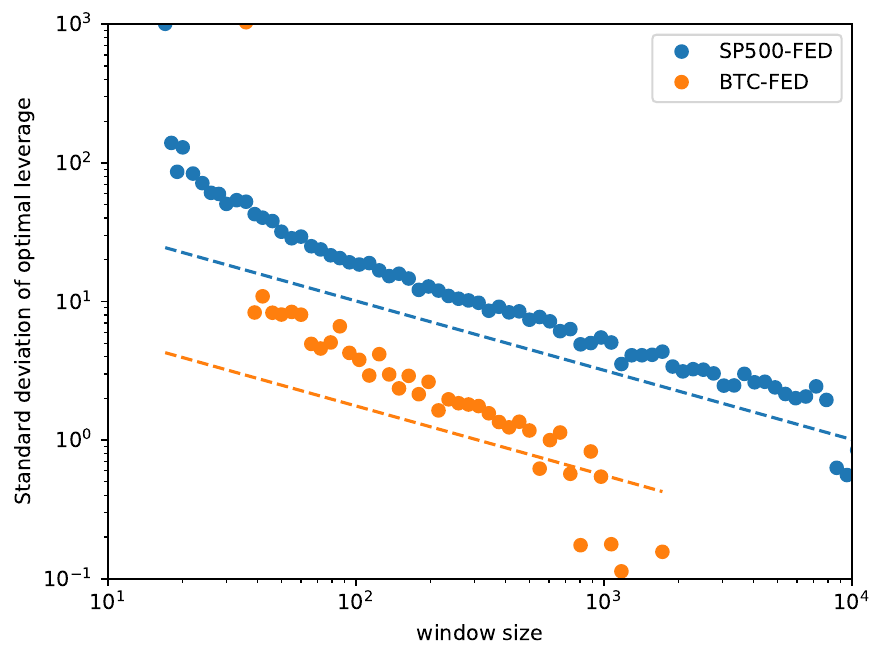}}
\end{picture}
\caption{Standard deviation of observed optimal leverage for \SPtick (red) and BTC (blue).\newline
The standard deviation of $\loptc$ in the simple case (symbols) as a function of window  length can be predicted based on the specific model \eref{lopt_sd} (straight lines), using the parameters found in section~\ref{TheEntire}. Only non-overlapping windows were used.}
\flabel{loglog}
\end{figure}

In \fref{l_opt_expanding} the diminishing fluctuations in $\loptc$ are
illustrated as follows: for every day the optimal leverage is computed
for the longest available window (always starting on the earliest available date) 
for the simple and complex cases. As time passes, the optimal leverage approaches $\loptc=1$, especially for the complex case, and -- as predicted -- much faster for BTC. 

\begin{figure}
\begin{picture}(300,130)(0,0)
\put(-10,0){\includegraphics[height=.2\textheight]{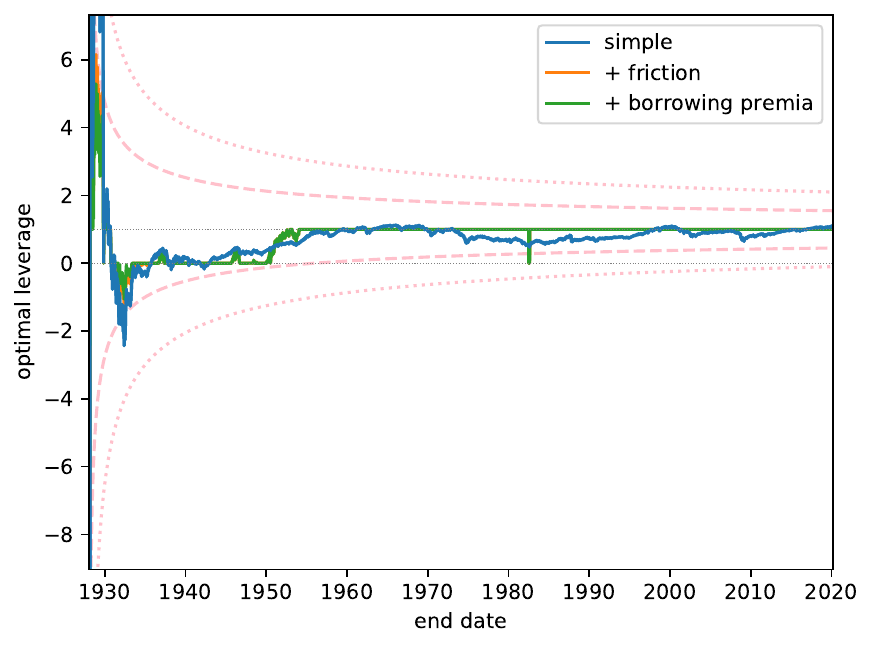}}
\put(180,0){\includegraphics[height=.2\textheight]{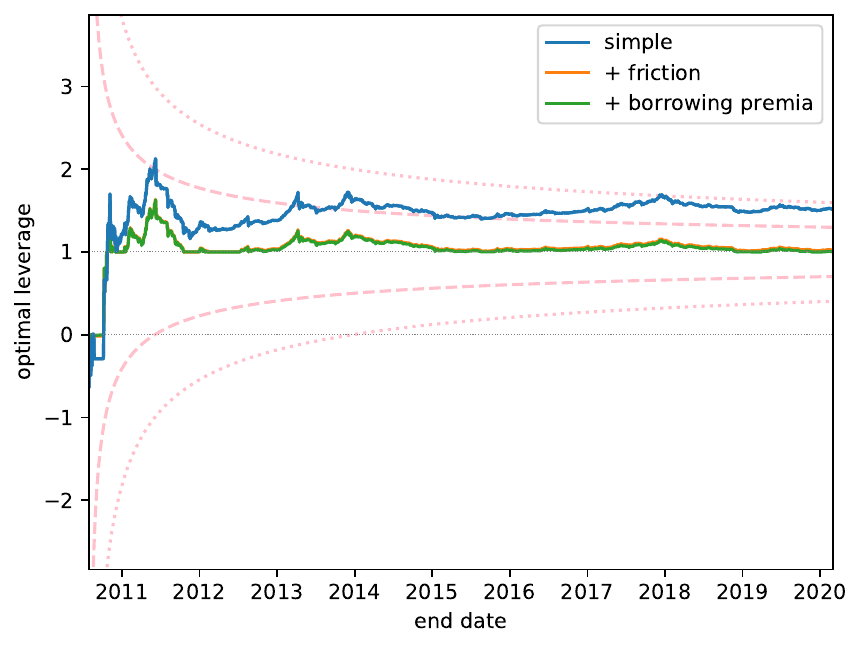}}
\put(50,130){a) \SP}
\put(260,130){b) Bitcoin}
\end{picture}
\caption{Daily observed optimal leverage for expanding windows.\newline
Also shown are the one- and two-standard deviation envelopes about $\loptc=1$, based on the estimates $\sigmac$ in section~\ref{TheEntire} for: a) \SP data beginning in 1927; and b) BTC data beginning in 2010.}
\flabel{l_opt_expanding}
\end{figure}

\Fref{l_opt_expanding} illustrates the convergence of $\loptc \to 1$ over time, but provides no information regarding the typicality of the time series.
Further insight into the dynamics of $\loptc$ can be gained by examining time series for fixed window lengths.
\Fref{l_opt_simple} shows $\loptc$ for the simple case for windows ranging from one to twenty years as a function of the end date of the window.
From the strong fluctuations over short time scales emerges attractive behavior consistent with the leverage efficiency hypothesis.
For the complex case (not shown) the effects of the stickiness of the points $\loptc=0$ and $\loptc=1$ are visible, and lend additional support to the hypothesis as it applies to real markets. 

\begin{figure}[h!]
\begin{picture}(300,140)(0,0)
\put(-10,0){\includegraphics[width=.51\textwidth]{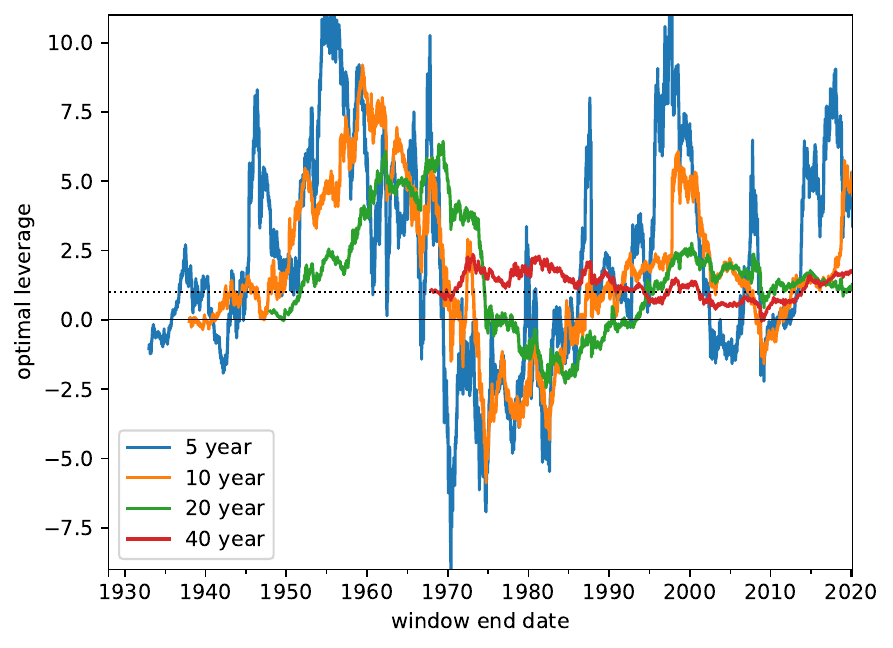}}
\put(180,0){\includegraphics[width=.498\textwidth]{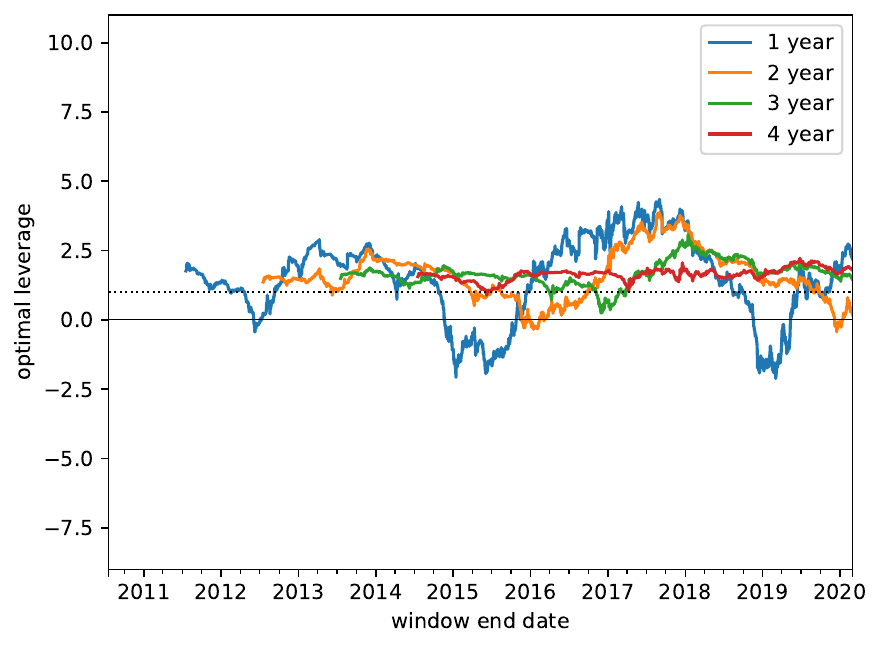}}
\put(50,130){a) \SP}
\put(260,130){b) Bitcoin}
\end{picture}
\caption{Observed optimal leverage for fixed-length windows.\newline
a) \SPtick b) BTC. Observed optimal leverage, $\loptc$, fluctuates strongly on
short time scales but appears to converge to $\loptc=1$ on long time scales,
which constitutes the central result of the study. The convergence for BTC is much faster (note the horizontal date scales).
}
\flabel{l_opt_simple}
\end{figure}
\FloatBarrier

\section{Discussion}\label{Discussion}

Nothing in nature, not even Brown's pollen \citep{Mazo2002}, truly
follows Brownian motion, whether geometric or not. Nor is anything in
nature knowably faithfully described by any mathematical expression
\citep{Renyi1967}. However, just as the movements of Brown's pollen, in
the appropriate regime, have some properties in common with Wiener
noise, so the movements of share prices have some properties in
common with geometric Brownian motion. Specifically, the daily excess
returns for the markets investigated -- like the fractional changes in geometric Brownian motion -- are sometimes positive and sometimes negative.  For any time window that includes both positive and negative daily excess returns, regardless of their distribution, a well-defined optimal
constant leverage exists in our computations, section~\ref{Simulations}. We have investigated empirically the properties of such optimal leverages.

Stability arguments, which do not depend on the specific distribution of returns and go beyond the model of geometric Brownian
motion, led us to the quantitative prediction that, on sufficiently long
time scales, real optimal leverage is attracted to $\loptr=1$.

We used specific properties of geometric Brownian motion to predict the fluctuations in $\loptc$,
providing a scale on which to judge our observations, which is supported by \fref{loglog}. 
While some parameter choices must be made, especially when attempting a degree of realism,
we consider our empirical findings a significant corroboration of our hypothesis. The points $\loptr=0$ and $\loptr=1$ are special due to the kinks in \fref{parabolic}. The economic paralysis argument suggests that $\loptr=1$ is a stronger attractor than $\loptr=0$, and our observations support this argument. One interpretation is that interest rates have been set at a level that strongly encourages investment in risky assets, so that, even in the presence of trading costs and possibly other constraints, investment is directed into business ventures rather than savings accounts.

Leverage efficiency also suggests a fundamental explanation for the existence of volatility in markets and, specifically, for its observed levels. Price fluctuations are necessary to avoid leverage instability and their observed amplitude is consistent with predictions that assume leverage efficiency. The corollary is that mainstream theories in which price fluctuations are caused by the public disclosure of information or by market dysfunction are, at best, incomplete. Trading at arbitrarily high frequencies will reveal structure, but this structure will be largely self-referential, without economic meaning beyond imposing market stability at ever shorter time scales. 

The existence of optimal leverage is important conceptually, and its observed value
and associated stability arguments are of practical
significance. While these arguments do not preclude special conditions
under which it is optimal to invest more than one's equity or to
short-sell an asset, they give a fundamental scale to leverage in
general. In other words, if it appears that optimal leverage is
outside the band $0 \leq \loptr \leq 1$, then a special reason -- such
as insider knowledge, a tax incentive, or a fraud -- for this violation of
leverage efficiency must exist. Artificially maintaining such
conditions will lead to instabilities. Consider housing: many
societies consider it desirable for an individual to be able to
purchase a home whose price exceeds his equity without having to take
reckless risks. Without carefully designed restrictions on speculative home
purchases, policies which aim to achieve the corresponding market
conditions, \ie $\loptr>1$, will defeat their purpose and create
investment bubbles followed by crashes or other unintended consequences.



Leverage efficiency is ``accountable'' in the sense of \citet[Chap.~I.2]{Popper1982}, who demanded that a ``theory will have to account for the
imprecision of the prediction''.\footnote{Popper does not refer to
  stochastic theories in this discussion. To apply his arguments to
  our case, specifically \eref{lopt_sd}, we replace ``precision in the initial conditions'' in his
  Chap.~I.3 by ``window length''. Both concepts quantify the
  information available about the system.} Leverage efficiency
predicts its own imprecision, \eref{lopt_sd}, and the degree of its
validity can be meaningfully and objectively tested. This is
particularly important given the complexity of the systems involved.

We emphasize that our work is in no way meant to advocate or evaluate
constant-leverage or any other investment strategies. Leverage
efficiency is a fundamental organizing principle for the stochastic
properties of markets. The data analysis in this study is an empirical
test of this fundamental principle. 

Our results are relevant to the equity premium, or price volatility, puzzle
\citep{MehraPrescott1985}. The values of volatility we observe in the most realistic analysis (the complex case)
are as predicted by excess return, \eref{noise}. In a simplistic analysis which ignores trading and other costs, we find volatility is {\it less} than one would expect, not {\it more} as consumption-based asset pricing models tend to suggest. We consider the puzzle essentially solved by our analysis.

Arguing in the context of the model, a strong
link between interest rates and leverage is \eref{lopt}: reducing the risk-free interest
rate $\murm$ (something we liken to the rate at which governments
lend) increases optimal leverage because, assuming that overall
drift $\mum$ does not change, it implicitly increases $\muem$
and creates an incentive to invest rather than save. This tends to
lead to an eventual increase in real leverage. Conversely, increasing the risk-free 
interest rate reduces optimal leverage.
We remark that effecting a change in real leverage, for instance in an attempt to avoid asset bubbles, through a change in $\murr$ is only one option available to policymakers. In some situations, the more direct approach of prohibiting leverages outside a range might be more appropriate.
The prediction of a range for the optimal leverage of a finite-time investment, \eref{lopt_range}, allows the detection of specific investments whose observed optimal leverages lie outside it. If the prediction is believed generally, then special circumstances must apply to such investments. One such circumstance is fraud: prices which do not arise from the trading feedbacks that give rise to leverage efficiency, for example because they are made up, will not have statistical properties consistent with leverage efficiency. We propose that such inconsistencies be used to detect fraudulent investment schemes and show in section~\ref{Tests} that the infamous Ponzi scheme of Bernard Madoff and associates might have been detected sooner, and without great difficulty, using this idea.

We do not consider our arguments specific to traditional financial markets, as demonstrated by their surprising applicability to BTC. They
are relevant also to other regularly traded assets and commodities,
related even to such basic needs as food and shelter.
They are relevant to
macroeconomic decisions. Indeed, the same type of dynamics --
multiplicative growth with fluctuations -- is at work in many other
systems. \Eref{motion} has been used to describe the growth of
populations in ecology \citep{LewontinCohen1969}, the early spread of
a disease in epidemiology \citep{DaleyGani1999}, and as the basis for the evolution of cooperation \citep{PetersAdamou2015a}.

We have argued that $0\leq \loptr\leq 1$ is a natural attractor for an
economic or market system, with the end points of the interval being
sticky. We note that a sticky $\loptr=0$ may correspond to a depression, in which there is no incentive to invest and to take risks. The aim
of economic policy may be viewed as creating conditions where $\loptr$
for the entire economy is close to one. 

\section*{Acknowledgements}
We thank Zonlab ltd.~for support, G.~D.~Nystrom for bringing the equity premium puzzle to our attention, I.~Lehti for pointing us to the \SP total return data set, A.~Messinger for suggesting we study Berkshire Hathaway returns, and C. Connaughton for invaluable help in preparing the codes for publication.

\bibliographystyle{abbrvnat}
\bibliography{./bibliography}

\end{document}